\documentclass[a4paper]{article}

\usepackage{19lomcon} 
\usepackage{cite} 
\usepackage{epsfig} 
\usepackage{epstopdf}

\begin{document}


\title{PARADIGMS AND SCENARIOS FOR THE DARK MATTER PHENOMENON  }

\author{Paolo Salucci \email{salucci@sissa.it}
 and
 Nicola Turini \email{turini@cern.ch}
}

\affiliation{$^a$ SISSA, Via Bonomea, 265, Trieste. INFN, QGSKY,  Sezione di Trieste}

\affiliation{$^b$ University of Siena, DSFTA Sezione di Fisica, Via Roma 56, Siena, 
INFN, Gruppo collegato di Siena, Siena}


\date{}
\maketitle

 
\vskip 1cm
\begin{abstract}

\vskip 0.25cm 

Well known scaling laws among the structural properties of the dark and the luminous matter in disc systems are too complex to be arisen by two inert components that just share the same gravitational field. 
This brings us to critically focus on the 30 year old paradigm, that, resting on 
a priori knowledge of the nature of dark matter (DM), has led us to a restricted number of scenarios, especially favouring the collisionless  $\Lambda$Cold Dark Matter one.
 
Motivated by such observational evidence, we propose to resolve the dark matter mystery by following a new paradigm: the nature of DM must be guessed/derived by deeply analyzing the properties of the dark and luminous mass distribution at galactic scales. The immediate application of this paradigm leads us to the existence of a direct interaction 
between dark and Standard Model particles which  has finely shaped the inner regions of galaxies. 
\end{abstract}

\vskip 0.25cm
 
\section{Introduction}
 
The mass distribution in Spirals is largely dominated 
by a dark component as it is evident from their kinematics and their other 
tracers of the mass distribution (e.g. see \cite{s19}).
The Dark Matter is thought to be made of particles that interact with 
Standard Model particles and with itself (almost) only via Gravitation. In the past 30 years people, in order to approach  the 'DM mystery', have adopted 
 the paradigm according to which one starts from a strong theoretical argument that
 leads us to a well defined and verifiable scenario and to a specific dark 
particle, detectable by experiments and astrophysical observations. This paradigm has pointed especially to the scenario of a stable Weakly Interacting Massive Particle (WIMP), likely coming from SuperSymmetric extensions of the Standard Model of elementary particles
\cite{Steigman_1985,Ber}. 

However, the above collisionless $\Lambda$CDM scenario has, at galactic scales, serious problems including that for which the predicted structural properties of DM halos result in strong disagreement with respect to those inferred from the internal motions of galaxies (see, e.g. \cite{s19}).
It has been claimed that these strong discrepancies can be eliminated by astrophysical processes (e.g. \cite{dicintio}),
however, as new data come in, the DM halos density profiles result always more difficult to be accounted for by such processes (e.g. cite{ks,dps}). Moreover, we  also have to consider that WIMP particles  have not turned up in any way, so far (see e.g. \cite{A19,Fr}).

 Thus, in order to successfully investigate the 'dark matter mystery' We believe that to propose a new paradigm is necessary.

 \section{Evidences calling for a change of paradigm}
 
 The structural components of normal Spirals include the well-known exponential thin disc, with surface density profile \cite{Freeman}:
\begin{figure*}
\center\includegraphics[width=12.cm]{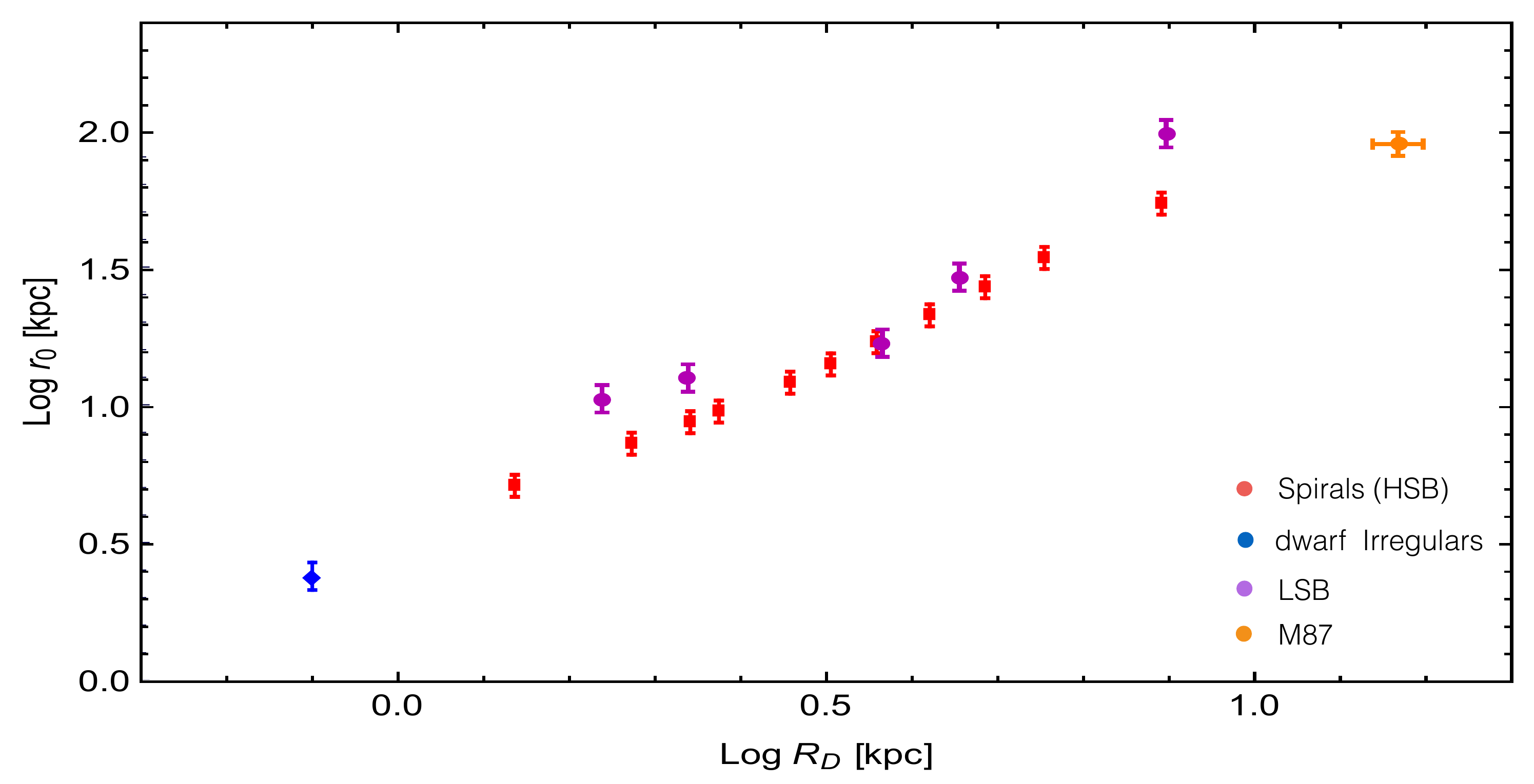}
\caption{ Log $r_0$ {\it vs} log $R_D$ in normal Spirals ({\it red}), dwarf Spirals ({\it blue}), Low Surface Brightness ({\it magenta}) and the giant elliptical M87 ({\it orange})}
\label{Fig1}
\end{figure*}

 \begin{equation}
 \label{eq1}
\mu (r;M_D) = \frac {M_D}{2 \pi R_D^2}\ e^{-r/R_D} 
\end{equation}
with $R_D$ the disc scale length that we can derive from galaxy photometry and $M_D$ the disc mass. 
We can derive $\rho_\star (r;M_D)$, the stellar volume density, by assuming that the disk has a thickness $0.1 \ R_D$, as found in edge-on Spirals, then: 
$$
\rho_\star(r; M_D)=\frac {\mu(r,M_D)} {0.1 R_D}
$$

 The DM halo component is assumed to follow the Burkert halo profile (\cite{SB}):
 \begin{equation}
 \label{eq_Burkert}
 \rho_{B}(r)=\frac{\rho_0 r_0^3}{(r+r_0)(r^2+r_0^2)} \quad 
 \end{equation}
 with $\rho_0$ the central density and $r_0$ the core radius. 
  
 The velocity model is:

 $$V_{mod}^2(r; r_0 ,\rho_0, M_D)=V_B^2(r; r_0, \rho_0) +V_D^2(r;M_D)
 $$
 
 with: $$ V_D^2(y,M_D)=\frac{G M_D}{2R_D} y^2 B\left(\frac{y}{2}\right)$$
 
 where $y\equiv r/R_D$, $G$ is the gravitational constant and $B=I_0K_0-I_1K_{1}$  is a combination of Bessel functions and with:  
 
 $$ V^2_B(r;r_0,\rho_0) =6.4 \ \frac{\rho_0r_0^3}{r}( \ln (1+\frac{r}{r_0})-\arctan \frac{r}{r_0}) +\frac{1}{2}\ln ( 1+\frac{r^2}{r_0^2}))
 $$.

 The model  has 3 free parameters that alongside with $R_D$, emerge all as specific functions of $M_{vir} $
 
 $$
 M_{vir}\equiv M_{DM}(R_{vir})= 4/3 \pi \ 100 \ 1 \times 10^{-29} \ R_{vir}^3
 $$
 see  Eqs. (6a)-(10) in \cite{s7}). The two densities take then the  form:
 
 $$\rho_{B}(r, r_0(M_{vir}),\rho_0(M_{vir}));\\\ \rho_{\star} (r, M_D(M_{vir}), R_D(M_{vir}))$$.

 The first amazing relationship that we find in Spirals features the size of the DM constant density region $r_0$ which is found to tightly correlate with the stellar disc scale length $R_D$ (see Fig. \ref{Fig1}). We have: 
 
 $$Log \ r_0= (1.38 \pm 0.15) \ Log \ R_D + 0.47 \pm 0.03 $$. 
 
 This relationship, first found in \cite{D}, is  confirmed today in 2300 Spirals \cite{lsd}, in 72/36 LSB and dwarf Irregulars and in the giant cD galaxy M87 \cite{dps,ks}. Overall, the relationship extends over three orders of magnitudes in galaxy luminosity. Noticeably, the quantities involved $r_0$ and $R_D$ are derived in totally independent ways: by accurate modelling of the galaxy kinematics and by using the galaxy photometry.

\begin{figure}[t]
 
\center\includegraphics[width=7cm]{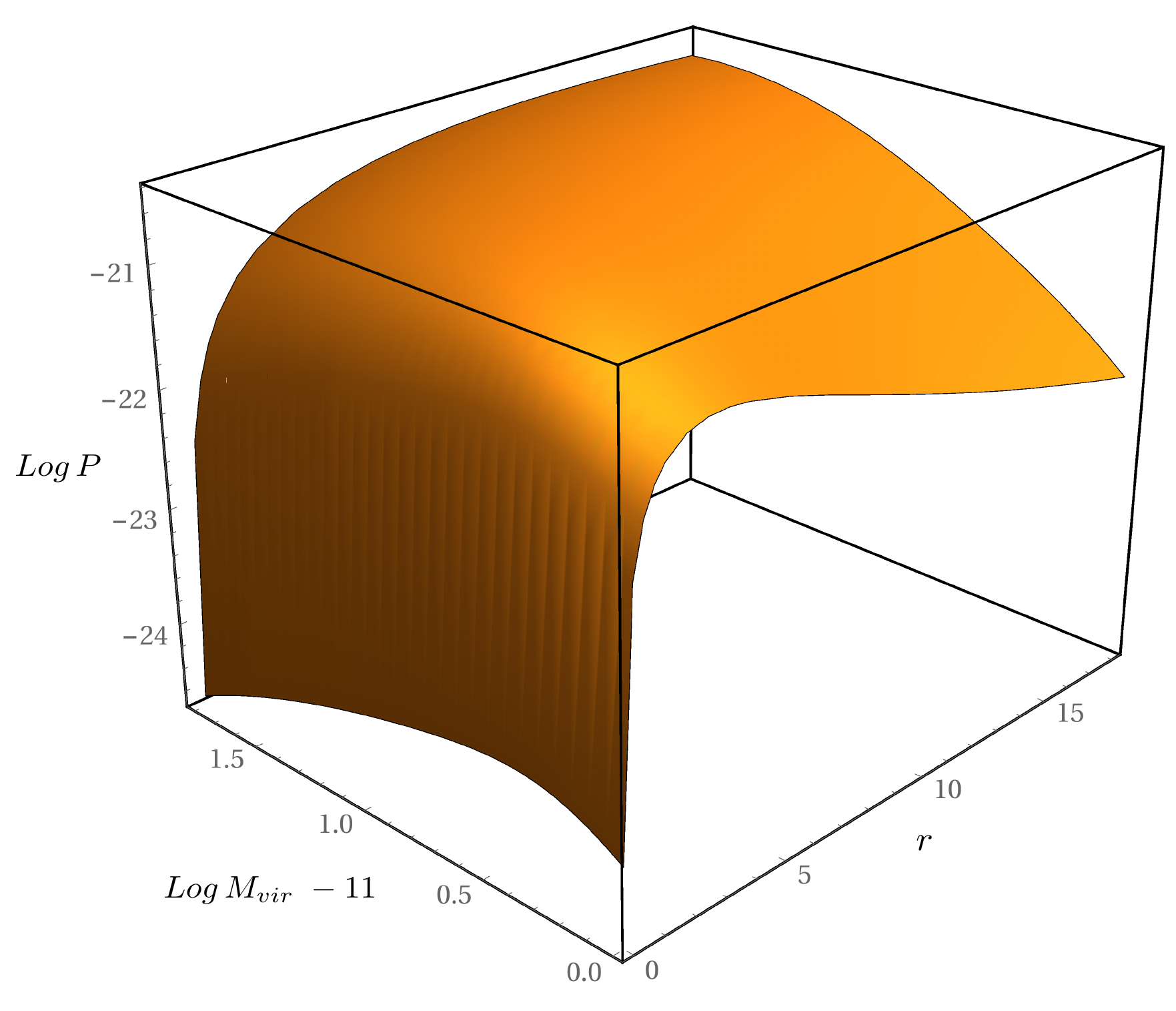}
\vskip 0.75cm
\center\includegraphics[width=6.0cm]{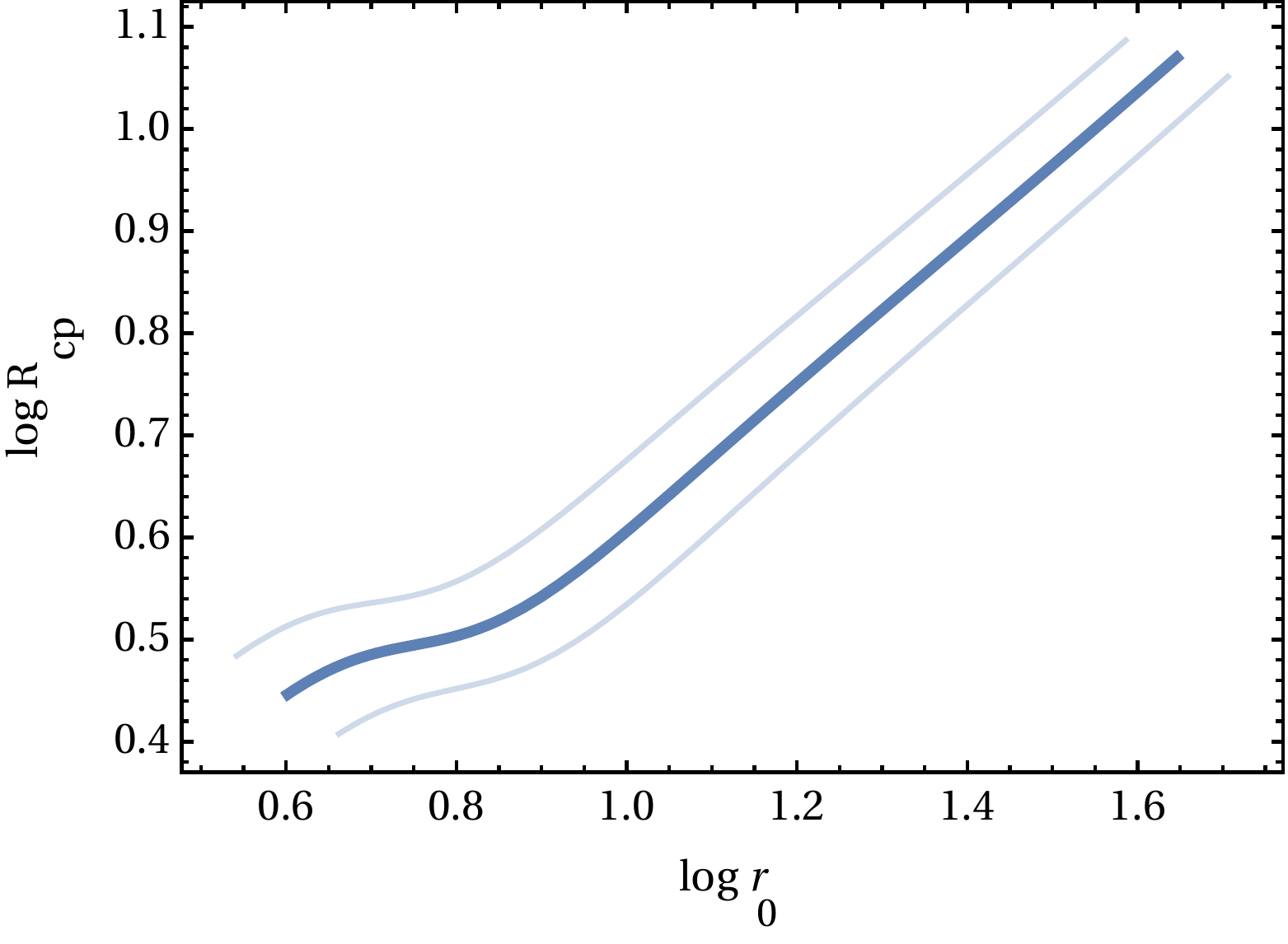}
 
\caption{ {\it Up.}  The DM pressure in Spirals (cgs units) as function of halo mass (in solar masses) and radius (in kpc). {\it Bottom.} Log $R_{cp}$ {\it vs.} Log \ $r_0$ }
\label{fig2}
\end{figure}

The second feature involves the fact that in Spirals, the luminous and the dark  central surface densities:

$$\Sigma_{0, \star} = \frac{M_D}{4 \pi R_D^2} \ \ \  \Sigma_{0,B} = \rho_0\, r_0$$

are found to strongly correlate (see \cite{ggn} and its Figs (1)-(3)), independently of whether the central region of a galaxy be DM or LM dominated. 
 
These two above relationships have no evident justification from currently accepted first principles. Let us assume for Spirals a spherical symmetry and that the DM halos are isotropically pressure supported; thus, the latter can be written in terms of the circular velocity $V(r)$ (\cite{Bi}): 

$$
P(r, M_{vir})= 1/3 \ \rho_{B}(r) V(r,M_{vir})^2
$$

with 
 $$V(r,M_{vir})= V_{mod}(r, r_0(M_{vir}), \rho_0(M_{vir},,  M_D(M_{vir})$$ 

and $Log ( M_{vir}/M_\odot) $ ranges from $10.9$ to $12.7$ (\cite{s7}). $P(r, M_{vir})$ (see Fig. \ref{fig2}) is null at the galaxy center, then, increases outwards reaching a maximum value at $r=R_{cp}$. We define the latter  as the ``constant pressure" radius where $dP/dr=0$ and finally strongly declines outwards.Since we have  that: $R_{cp} \simeq r_0$ (see Fig. \ref{fig2}) hereafter, we will consider these two quantities as the same. 
 
Remarkably, $P(r_0(M_{vir}),M_{vir})$ varies less than a factor 1.5 among Spirals, supporting the interpretation of the radius $r_0$ as the edge of the region where the DM-LM interactions have taken place so far. 
In analogy with the self-annihilating DM case in which the density kernel is:
$K_{SA}(r)= \rho_{DM}^2(r) $ we define $K_C(r)$ as the density kernel of the DM-baryons interaction:
\begin{equation}
 \label{eq13}
K_C(r)=\rho_{DM}(r)^a \rho_\star(r)^b \ v^c 
\end{equation}

where $ v$ is the relative    velocity between dark and Standard Model particles. 
The exact form for $K_C(r)$ is unknown, however, definiteness and simplicity suggest us to assume: $a=1$, $b=1$, $c=0$. Let us stress that the kernel $K_C$ is defined at a macroscopic scale, i.e., it is spatially averaged over a scale of the order of the variations of the galaxy gravitational field which is in the range of  1-10 kpc. On a microscopical level, where the interactions  really take place, the interacting kernel could be much more complex, variable and strongly depending on the relative velocity. 
 
\begin{figure}[t]
\center\includegraphics[width=6.5cm]{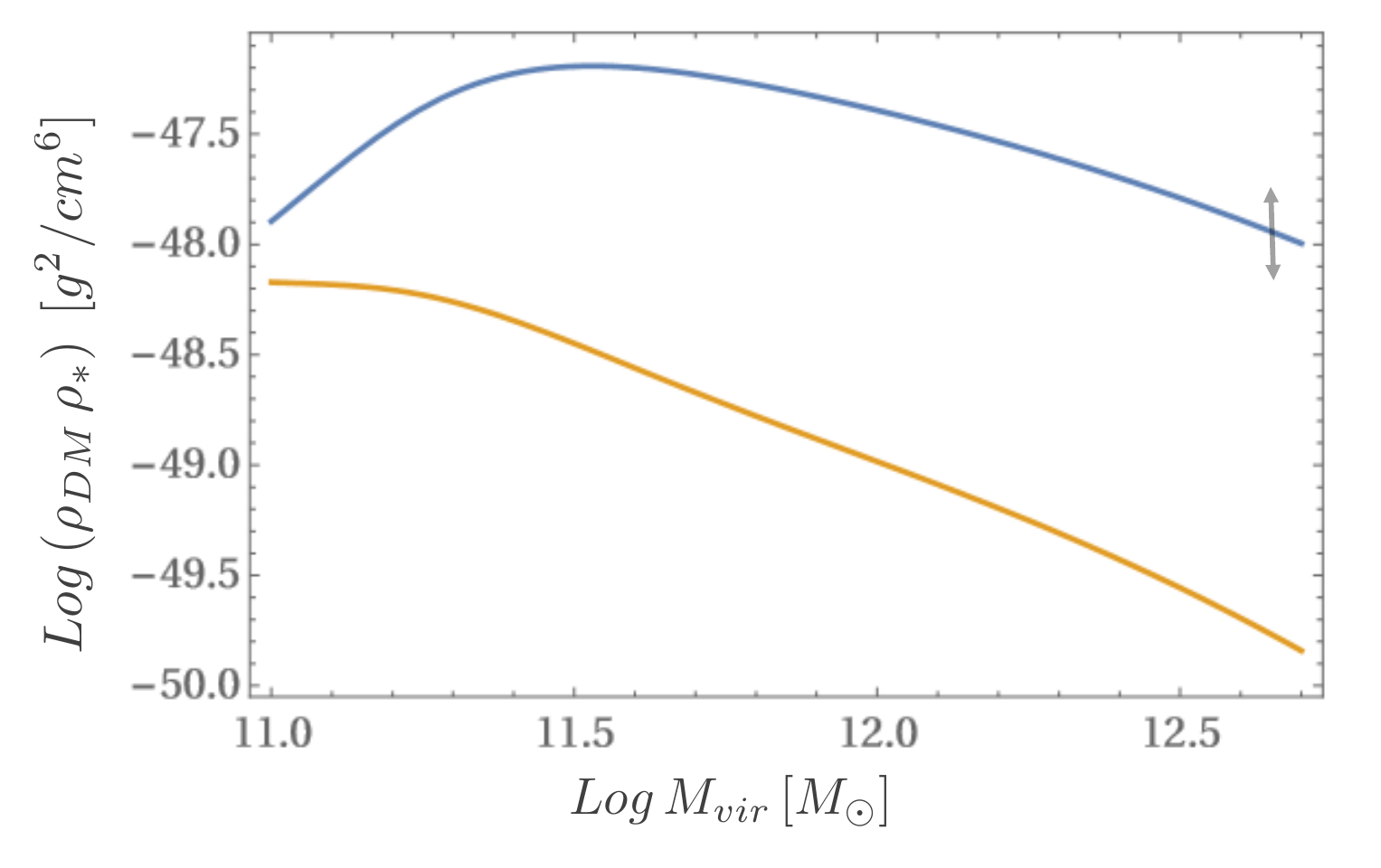}
\center\includegraphics[width=8.5cm]{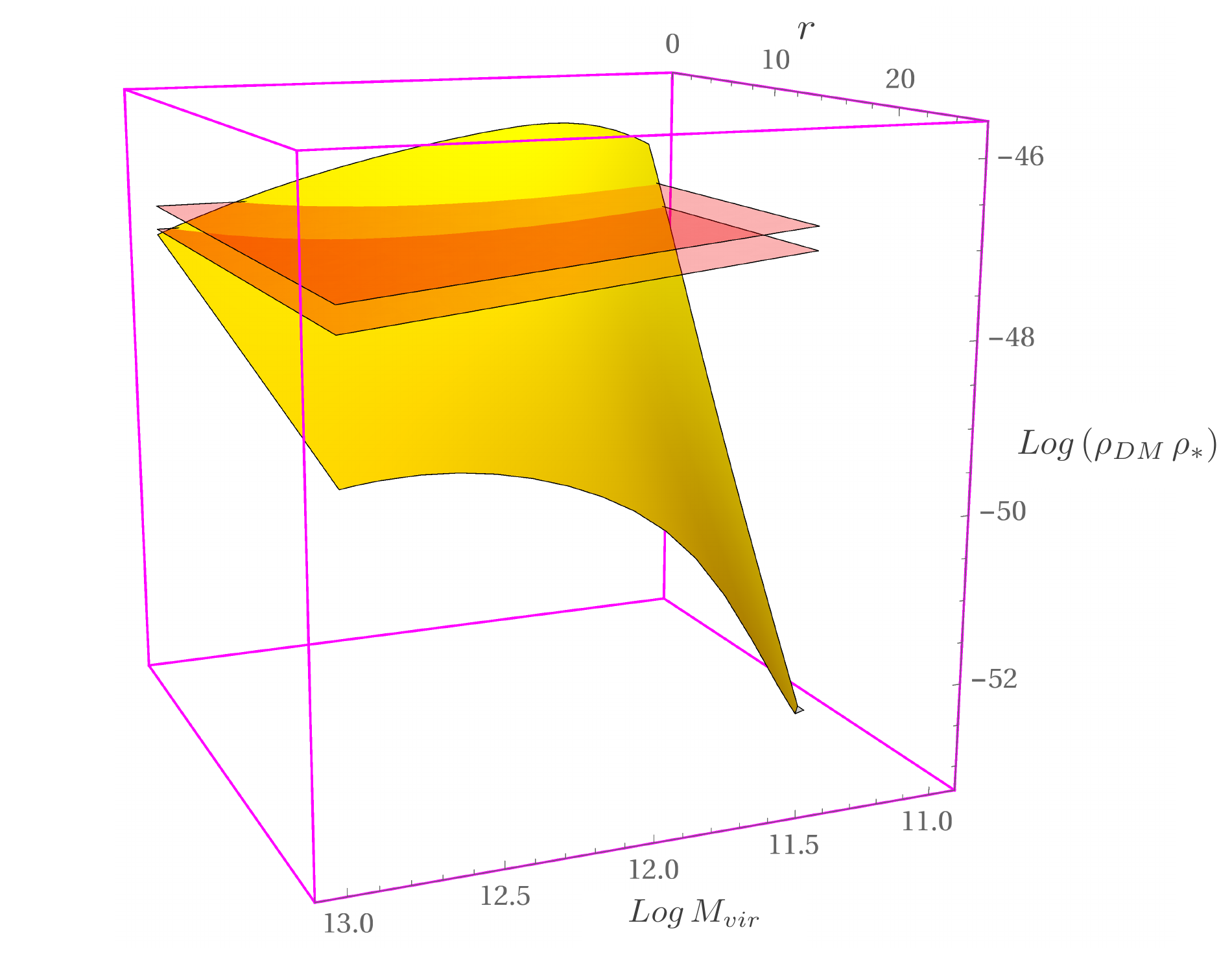}

\caption{ {\it Up.} $log \ K_C(r_0)$   as a  function of $log \ M_{vir}$ ({\it blue line}). The run of $K_{SA}(r_0)$ is also shown ({\it orange line}). {\it Bottom.} log $K_C/ \,(g^2cm^{-6})$ as function of log $ M_{vir}/ M_\odot$ and $r/kpc$ ({\it yellow surface}). In Spirals, the full range of $K_C(r_0)$ lies between the two parallel planes.}

\label{Fig4}
\end{figure}

We evaluate in Spirals $K_C(r_0(M_{vir}),M_{vir})$ and we find that: 
\begin{equation}
 \label{eq14}
K_C(r_0) \simeq \ const
=10^{-47.5 \pm 0.3} g^2 cm^{-6} \quad .
\end{equation}
 the above kernel keeps constant within a factor of about 2, see Fig. \ref{Fig4}.
In comparison, in the same objects and at the same radii, $K_{SA}(r_0)$
varies by two order of magnitudes. It is also impressive in Fig. \ref{Fig4})  that $ K_C(r,M_{vir})$ varies largely both among galaxies and in each galaxy, but, at $r\simeq r_0$,
 always takes that  value above,  which  marks the edge of the  sphere  inside which the dark-luminous matter interactions 
have occurred so far. 

We realize that some non gravitational energy could have been directly exchanged between atoms (or photons and/or neutrinos) and DM particles via processes currently unknown and certainly challenging the presently agreed first principles of Physics. 

The  DM-LM entanglement  in galaxies presented  in this section   works as a strong motivation for advocating a new Paradigm, according to which, the Nature of the dark particle and its related Scenario have to be determined  from reverse-engineering the galactic observations that  definine the DM phenomenon. 

\begin{figure}[t]
\center\includegraphics[width=11.5cm]{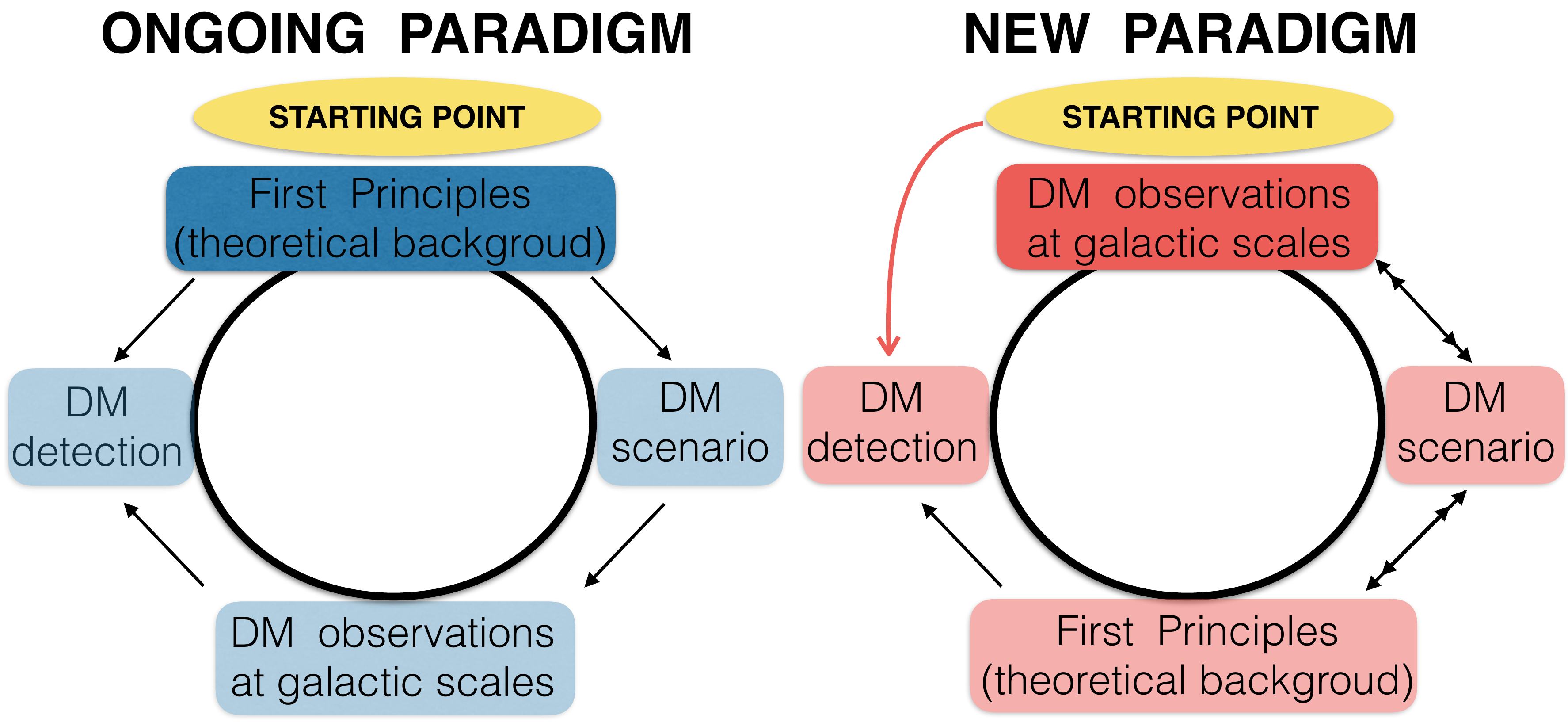}
\caption{The current and the new Paradigms. Notice the different role of the galactic DM-related observations. }
\label{npar}
\end{figure}

 Our  Paradigm has the following loop, see Fig.\ref{npar}: observations lead us to a new DM scenario that, once verified by other purposely planned observations, will provide us with the theoretical background of the DM phenomenon.

\section{The Interacting DM Scenario}

 By applying the new paradigm, a scenario immediately arises: it features dark-luminous matter interactions relevant when summed up to the age of the Universe  $\simeq 10^{10} \,yrs$ and occurring in very dense regions of dark {\it and} luminous matter. On the other hand, on the galaxy free-fall time, at high redshifts and in the outermost halo's regions, the DM particle behaves in a collisionless way. 

The particle itself is presently unspecified: our scenario is open to a huge field of possibilities that should be followed up  by suitable observations and experiments. However, the DM particle - nucleon interactions have left behind, in galaxies, a number of imprints, including the existence of DM density cores. 

\begin{figure}[t]
\center\includegraphics[width=7.cm]{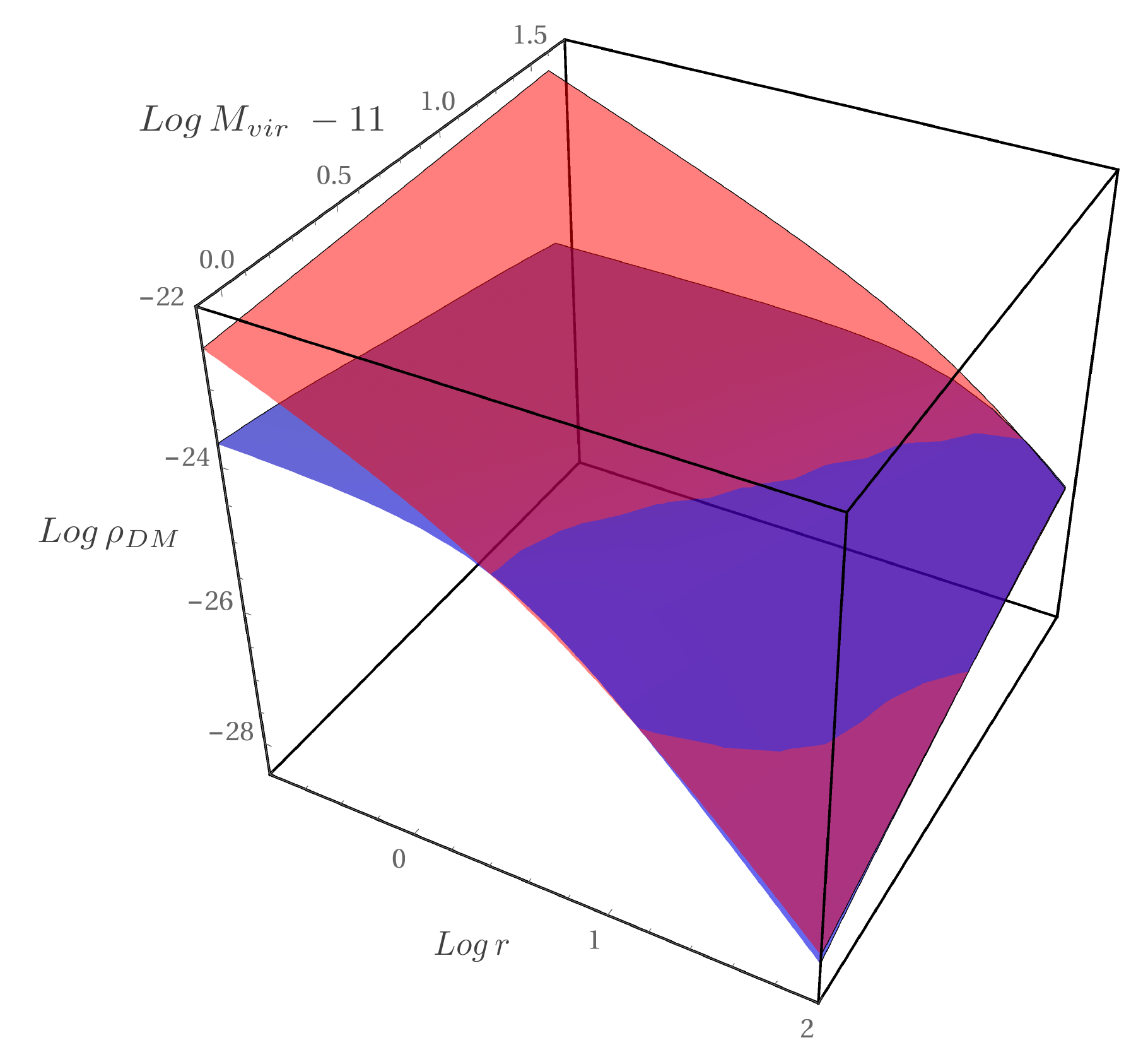}
\center\includegraphics[width=6.cm]{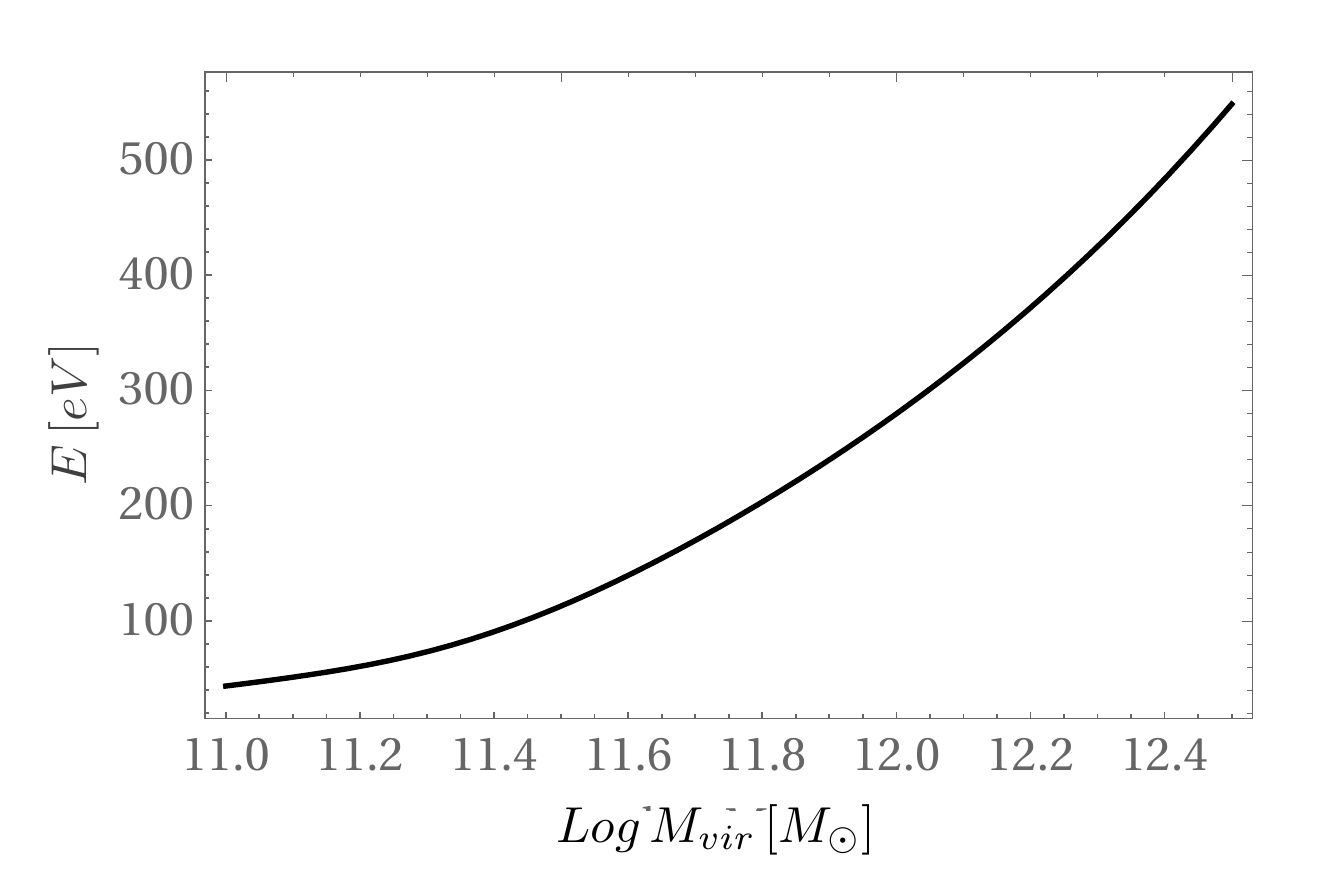}
 
\caption{ {\it Up.} Primordial ({\it red}) and present-day ({\it blue})  dark matter density profiles. $log \ \rho_{DM}$  in  $  g \, cm^{-3}$ is shown as a function of log radius (kpc) and log halo mass (in solar masses). {\it Bottom.} The energy of a core-forming SM-DM particles interaction  for a GeV of the particle mass as function of log $M_{vir}$.}
\label{Fig6}
\end{figure}

In our scenario, dark halos were formed with the NFW density profile \cite{nfw} which is characteristic for the collisionless particles. Remarkably, this profile is  recovered in the outermost regions of the present day galactic halos of Spirals: i.e.  for $r>2 r_0  $ (\cite{s7}), where we find $\rho_{DM}=\rho_{NFW}$ (see Fig. \ref{Fig6}a), with: 

\begin{equation}
\rho_{NFW}(r,M_{vir})= \frac{M_{vir}}{4\pi R_{vir}} 
\frac{c^2 g(c)}{\tilde{x}(1+c\tilde{x})^2} \quad, 
\label{NFW_density_profile}
\end{equation}
where $R_{vir}$ is the virial radius, $\tilde{x} =r/R_{vir}$, $M_{vir}$ is the viraial mass,

$$
c \simeq 14 \ (M_{vir}/(10^{11}M_\odot))^{-0.13}
$$ 

is the concentration parameter and $g(c)=[ln(1+c)-c/(1+c)]^{-1}$ (see \cite{s7}). This is  extraordinary since in the inner regions of Spirals actual DM profiles are in total disagreement with the NFW one. We, therefore, derive the primordial DM halo density by extrapolating the RHS of Eq. \ref{NFW_density_profile} down to $r=0$. Then we can derive, for an object of mass $M_{vir}$, the amount 

$$
\Delta M_{DM}= 4 \pi \int^{r_0}_0 (\rho_{NFW}(r, M_{vir}) -\rho_B(r,M_{vir})) r^2 dr 
$$

of DM removed inside $r_0$ by the core forming collisions.  Inside $r_0$ this amount is from $40\ \% $ to $ 90\ \%$ of the primordial mass  only $1\%$ of the (present) total halo mass.

Given $m_p$ the dark particle mass, the number of interactions per galaxy involved in the core-forming process is: 

$$
N_I(M_{vir})=\Delta M (M_{vir})/m_p
$$ 

The 
number of interactions for galaxy atom of mass $m_H$ is 

$$N_{I/A} = \frac{\Delta M(M_{vir})}{M_\star} \,m_H/$$. 

$W$, the work done  during the core-forming process is obtained  by $\frac {W}{4 \pi}$=
 
 $$  \int_0^{r_0}\rho_{NFW}(r;M_{vir}) M_{NFW}(r;M_{vir})\,r \, dr -  \int_0^{r_0} \rho_{B}(r;M_{vir}) M_B(r;M_{vir})\,  r\, dr $$
 
 We divide this energy by the number of interactions $N_I(M_{vir})$ taken place in each galaxy inside $r_0$ during the Hubble time and we get (see Fig. \ref{Fig6}b) the energy per interaction per GeV mass of the dark particle: 
 $$E_{core}=(100-500) \ eV \ \frac {m_P}{GeV} $$.
 
At a microscopic level, where do such interactions take place? The possibilities include gravitationally bound objects like planets, main sequence or giant stars or places with high baryonic density/temperature/velocity as white dwarfs, neutron and binary neutron stars, accretion discs to galactic black holes. 

Noticeably, in galaxies almost all the above locations have a radial distribution proportional to that of  the stellar disc. 

The DM halo particles, once inside $r_0$, while traveling through these locations a) acquire from collisions with the atoms an extra kinetic energy $E_{core}$ sufficient to leave the region or b) lose  an amount $\sim E_{core}$  of their kinetic energy by collisions or absorptions and are captured becoming  disc particles.

Alternatively, the presence of dense objects, like stars and BH, could enhance the DM self annihilation due to local DM density increase, depleting, in Hubble time scale, the central region of  galaxies of DM and creating, as in the previous cases, density cores strongly tied to the LM  distribution.

\section{Conclusion}

The Dark Matter Phenomenon features strong correlations between quantities deeply-rooted in the luminous world and quantities of the dark worlds. These relationships unlikely arise from some known first principle or as result of some known astrophysical process.

This has lead us to propose 
a new {\it paradigm} for the DM phenomenon, according to which the scenario for this elusive component should be obtained from reverse-engineering the DM-related observations at galactic scales, i.e,. a new frame of mind on the role of the inferred DM properties at such scales. 

By following this strategy we found that the quantity $\rho_B(r) \rho_{D}(r)$, which is the kernel associated to the SM-DM particles interaction, assumes, at $r_0$, the edge of the constant density region, almost the same value in all galaxies. This opens the way for a scenario featuring an interacting DM particle with a core-forming exchange of energy shaping the structure of the inner parts of the galaxy dark halos. 

We expect that this particle will show up from anomalies in the internal properties of the above locations (e.g. stars) On larger scales the interaction could radiate diffuse energetic photons detectable by VHE gamma rays experiments. The Moon and the Earth atmosphere may also be source of DM generated radiation. 
\vskip 2.5cm
 
{\bf Acknowledgments }
\vskip 0.5cm
We thank Chiara di Paolo for useful discussions. PS thanks Alexander I.Studenikin for the organization of the Lomonosov Conferences source of new ideas.

\end{document}